\documentclass[12pt,letterpaper,12pt,sort&compress]{article}
\usepackage[T1]{fontenc}
\usepackage[latin9]{inputenc}
\usepackage{amsmath}
\usepackage{amssymb}
\usepackage{graphicx}
\usepackage[numbers]{natbib}

\makeatletter

\providecommand{\tabularnewline}{\\}

\newcommand{\lyxaddress}[1]{
\par {\raggedright #1
\vspace{1.4em}
\noindent\par}
}

\@ifundefined{date}{}{\date{}}

\usepackage{indentfirst}
\usepackage{amsfonts}
\usepackage[T1]{fontenc}
\usepackage{ae,aecompl}
\usepackage{sidecap}
\usepackage[section]{placeins}
\usepackage{epsf}

\usepackage{fancyhdr}

\makeatother

\begin{document}

\title{\textbf{An atomistic view of grain boundary diffusion}}

\author{\textsf{Y. Mishin}}

\maketitle

\lyxaddress{\noindent \begin{center}
\vspace{-1cm}
\textsf{Department of Physics and Astronomy, MSN 3F3, George Mason
University, Fairfax, Virginia 22030, USA}\\
\textsf{Email: ymishin@gmu.edu}
\par\end{center}}

\noindent \textbf{Keywords:} Grain boundary, diffusion, modeling,
simulations, molecular dynamics, correlations, liquid.\bigskip{}

\noindent \textbf{Abstract.} This paper presents an overview of recent
computer simulations of grain boundary (GB) diffusion focusing on
atomistic understanding of diffusion mechanisms. At low temperatures
when GB structure is ordered, diffusion is mediated by point defects
inducing collective jumps of several atoms forming a chain. At high
temperatures when GB structure becomes highly disordered, the diffusion
process can be analyzed by statistical methods developed earlier for
supercooled liquids and glasses. Previous atomistic simulations reported
in the literature as well as the new simulations presented in this
paper reveal a close similarity between diffusion mechanisms in GBs
and in supercooled liquids. GB diffusion at high temperatures is dominated
by collective displacements of atomic groups (clusters), many of which
have one-dimensional geometries similar to strings. The recent progress
in this field motivates future extensions of atomistic simulations
to diffusion in alloy GBs, particularly in glass-forming systems.

\section{Introduction\label{sec:Introduction}}

It has long been known that atoms diffuse in grain boundaries (GBs)
orders of magnitude faster than in the crystalline grains, an effect
which is often referred to as ``short-circuit diffusion'' \citep{Kaur95}.
Fast GB diffusion can control many processes in materials, including
microstructure evolution, phase transformations, and certain modes
of plastic deformation and fracture \citep{Kaur95}. Over the past
decades, significant progress has been achieved in the accuracy and
reliability of GB diffusion measurements \citep{Kaur95,Mishin97e,Mishin99f}.
At the same time, many fundamental aspects of GB diffusion remain
poorly understood, including our knowledge of diffusion mechanisms
on the atomic level. The basic reason for this disparity lies in the
fact that the interpretation of experimental measurements is mostly
based on phenomenological models of diffusion such as the Fisher model
\citep{Fisher} and its numerous modifications \citep{Kaur95,Mishin97e,Mishin99f},
which in turn rely on the empirical Fick's law of diffusion. As a
result, experimental measurements deliver at best a numerical value
of the GB diffusion coefficient (often only its product with the GB
width). While important information for applications, these numbers
represent a convolution of many atomic-level factors and atomic events
averaged over the macroscopic time scale of diffusion measurements
(usually minutes to hours) and relatively large length scales (e.g.,
$\sim$ tens of micrometers) that greatly exceed the atomic scale.
Such measurements do not permit the extraction of reliable information
related to diffusion mechanisms. While diffusion in bulk solids has
been probed by other methods providing access to diffusive events
on the atomic scale \citep{Mehrer2007}, such as quasi-elastic neutron
scattering \citep{Vogl05,Springer05} or perturbed angular correlations
\citep{Collins2009}, applications of these methods to GBs are virtually
non-existent, and would be extremely difficult to implement given
the complexity of the GB structures and the relatively small fraction
of atoms residing in GB regions.

Given these difficulties, atomistic computer simulations offer the
only feasible way of getting at least a glimpse into the atomic-level
picture of GB diffusion. The goal of this paper is to review some
of the recent progress in this field. We start by surveying the existing
methods for the calculation of GB diffusion coefficients and investigation
of diffusion mechanisms in atomically ordered GBs at low temperatures.
We then shift the focus to diffusion in disordered GB structures arising
at high temperatures. Recent simulations results will be presented,
which reveal a striking similarity between the atomic motion in high-temperature
GBs, on one hand, and in the supercooled liquid phase of the same
material, on the other hand.

\section{Methods of GB diffusion calculations\label{sec:Methods}}

A number of simulation methods have been developed for computational
prediction of GB diffusion coefficients and investigation of GB diffusion
mechanisms \citep{Mishin2010a}. The most efficient methods utilize
molecular dynamics (MD) simulations with many-body semi-empirical
interatomic potentials \citep{Suzuki03a,Suzuki03b,Suzuki05a,Pun09a,Frolov09c,Frolov2013a}.
For metals and metallic alloys, embedded-atom method (EAM) potentials
\citep{Daw84,Mishin.HMM} are usually employed, while other systems
require more complex functional forms such as the modified embedded-atom
method (MEAM) \citep{Baskes87} or the angular-dependent potential
(ADP) \citep{Mishin2010a}. A GB of interest is created by standard
geometric constructions and thermally equilibrated at the desired
temperature. A long MD simulation is performed in which the atoms
can randomly diffusive along the boundary. At high temperatures, mean-squared
displacements of atoms within the GB core region are computed and
the GB diffusion coefficient $D_{gb}$ is extracted from the Einstein
relation $\left\langle x^{2}\right\rangle =2D_{gb}t$ for different
crystallographic directions $x$ within the GB plane. To collect adequate
statistics of atomic displacements, the simulations are conducted
at relatively high temperatures, typically above $0.6$-$0.7T_{m}$
($T_{m}$ being the melting point). In many cases, results of such
calculations demonstrate very good agreement with experimental data.
For example, the recently computed GB self-diffusion coefficients
in Cu \citep{Suzuki05a,Frolov09c} are in excellent agreement with
experimental measurements with radioactive isotopes \citep{Surholt97a}.
Good agreement was also found between computed and measured impurity
diffusion coefficients of Ag in $\Sigma5$ tilt GBs in Cu \citep{Frolov2013a}.
These examples demonstrate that due to the recent progress in this
field, atomistic simulations are now capable of predicting GB diffusivities
in metallic systems on the quantitative level. 

\begin{figure}
\noindent \negmedspace{}\negmedspace{}\negmedspace{}\includegraphics[clip,scale=0.033]{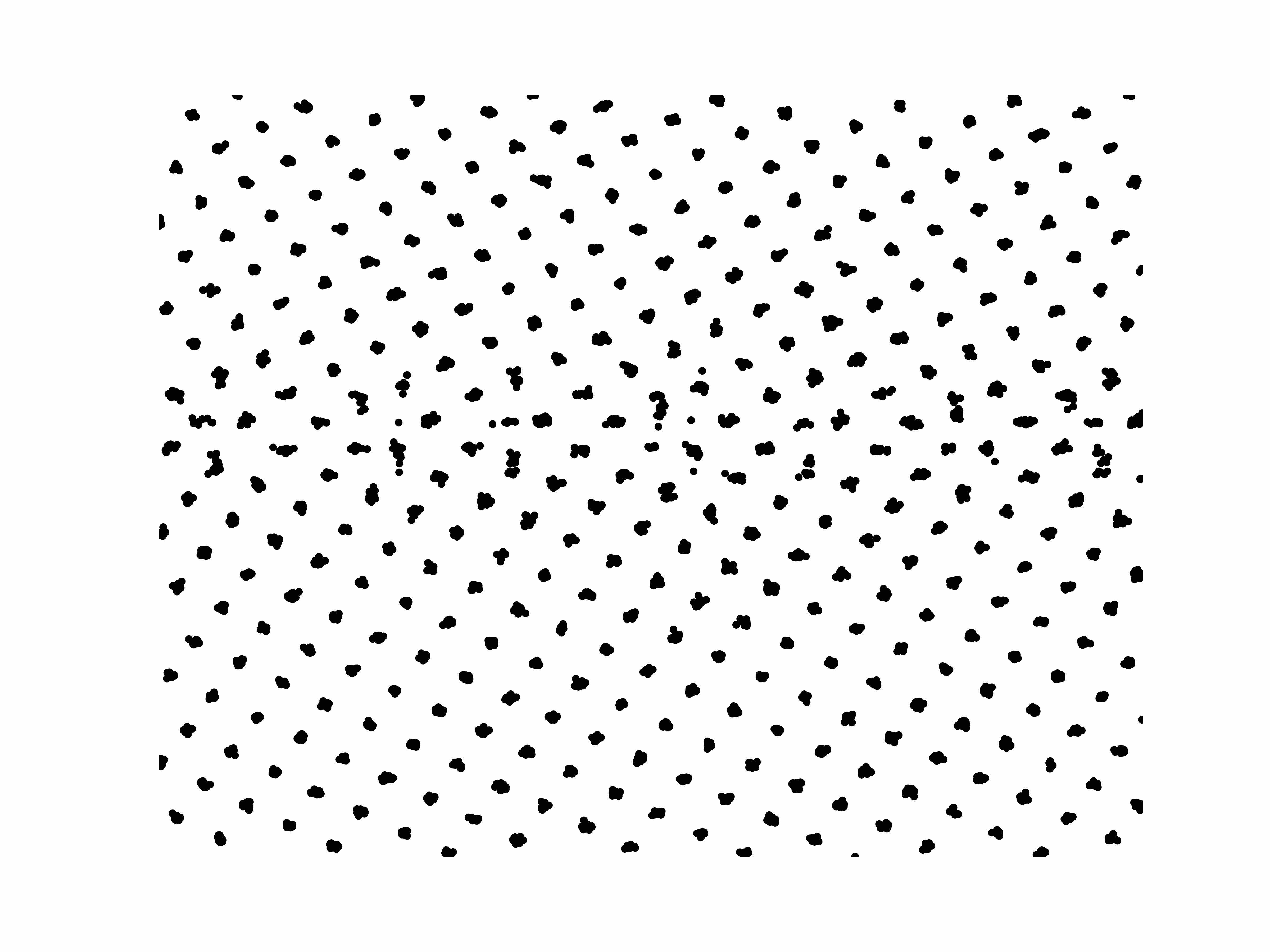}\negmedspace{}\negmedspace{}\negmedspace{}\includegraphics[clip,scale=0.033]{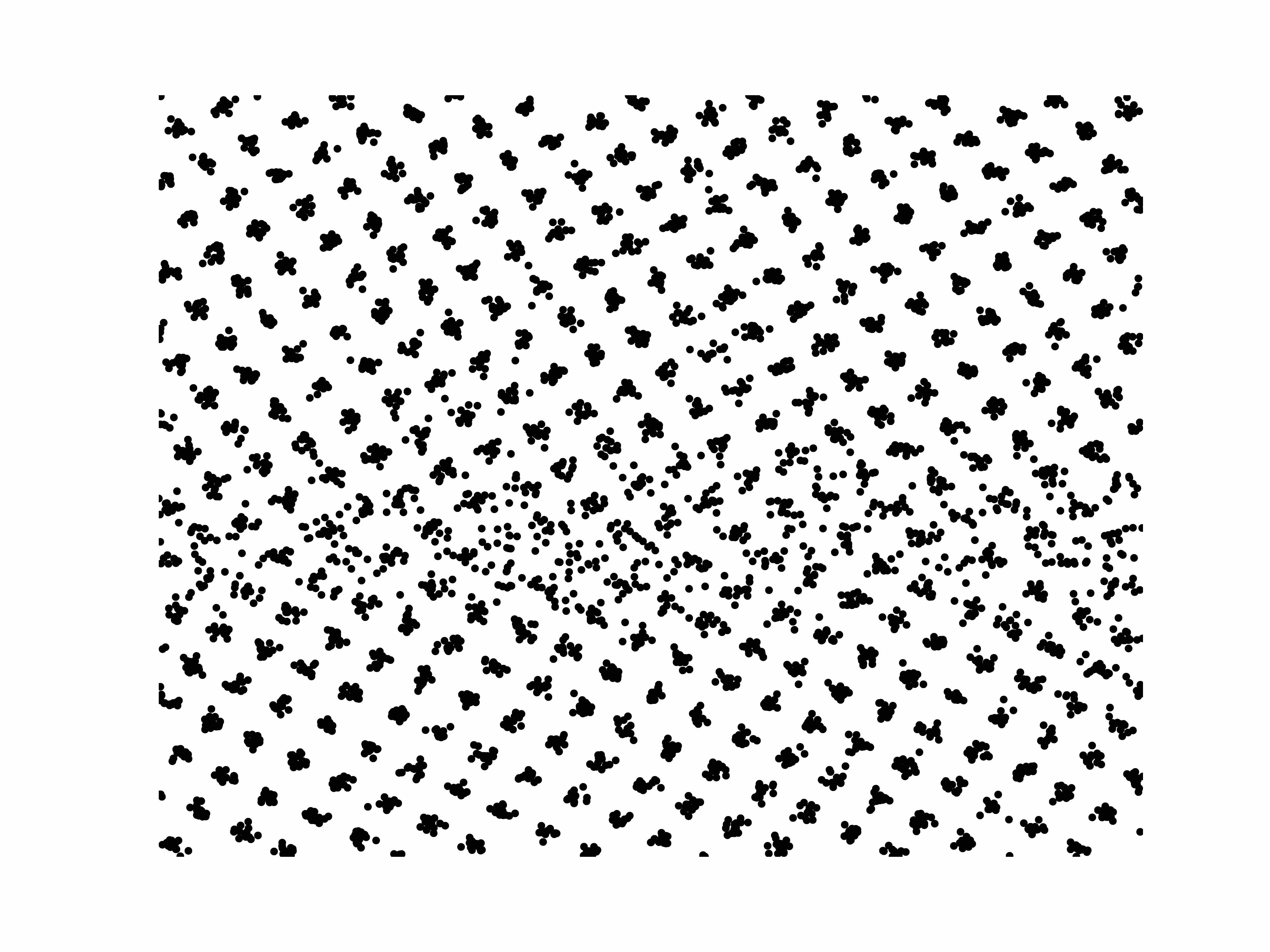}\negmedspace{}\negmedspace{}\negmedspace{}\includegraphics[clip,scale=0.033]{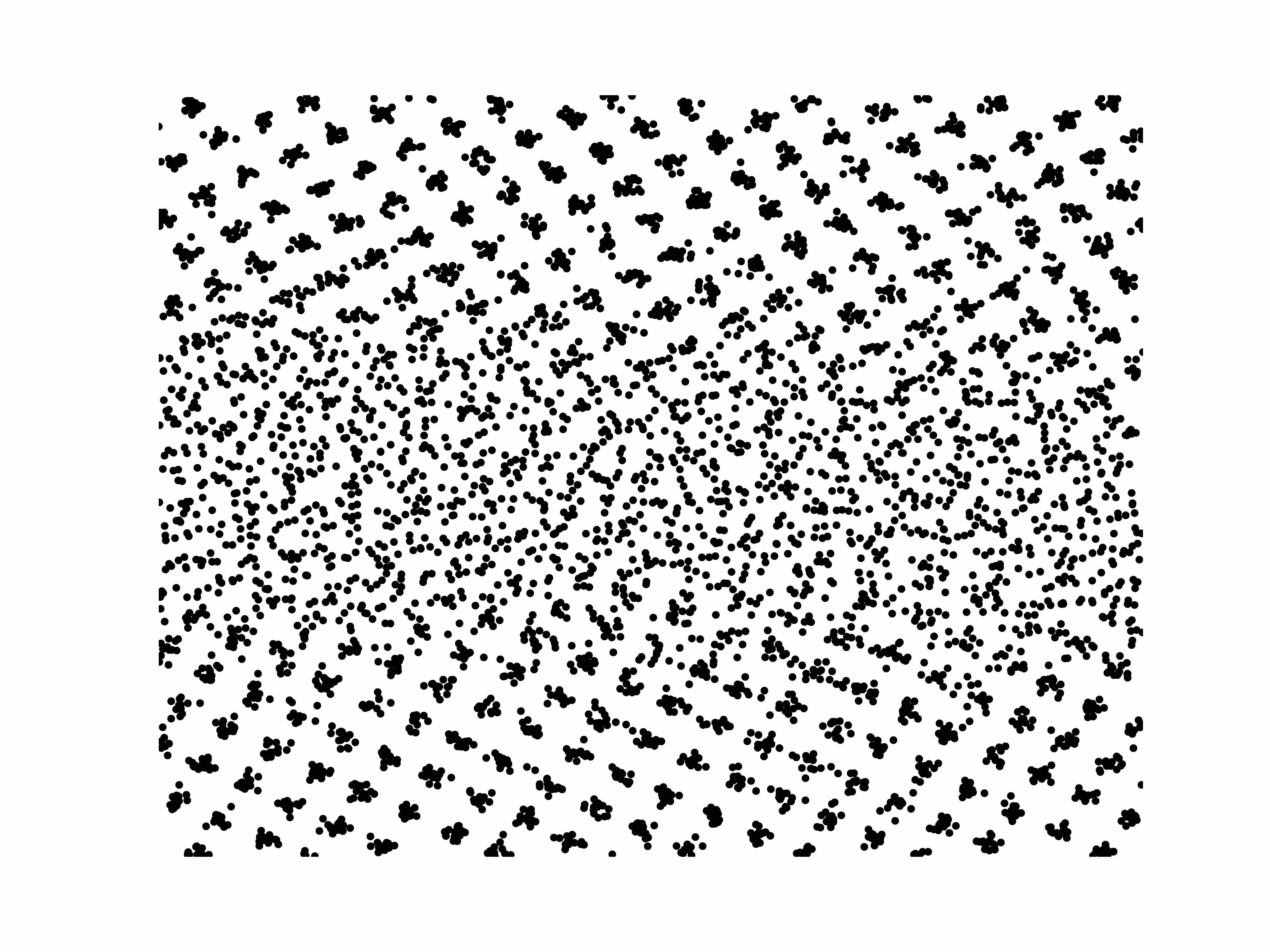}

\noindent \negmedspace{}\negmedspace{}\negmedspace{}\includegraphics[clip,scale=0.033]{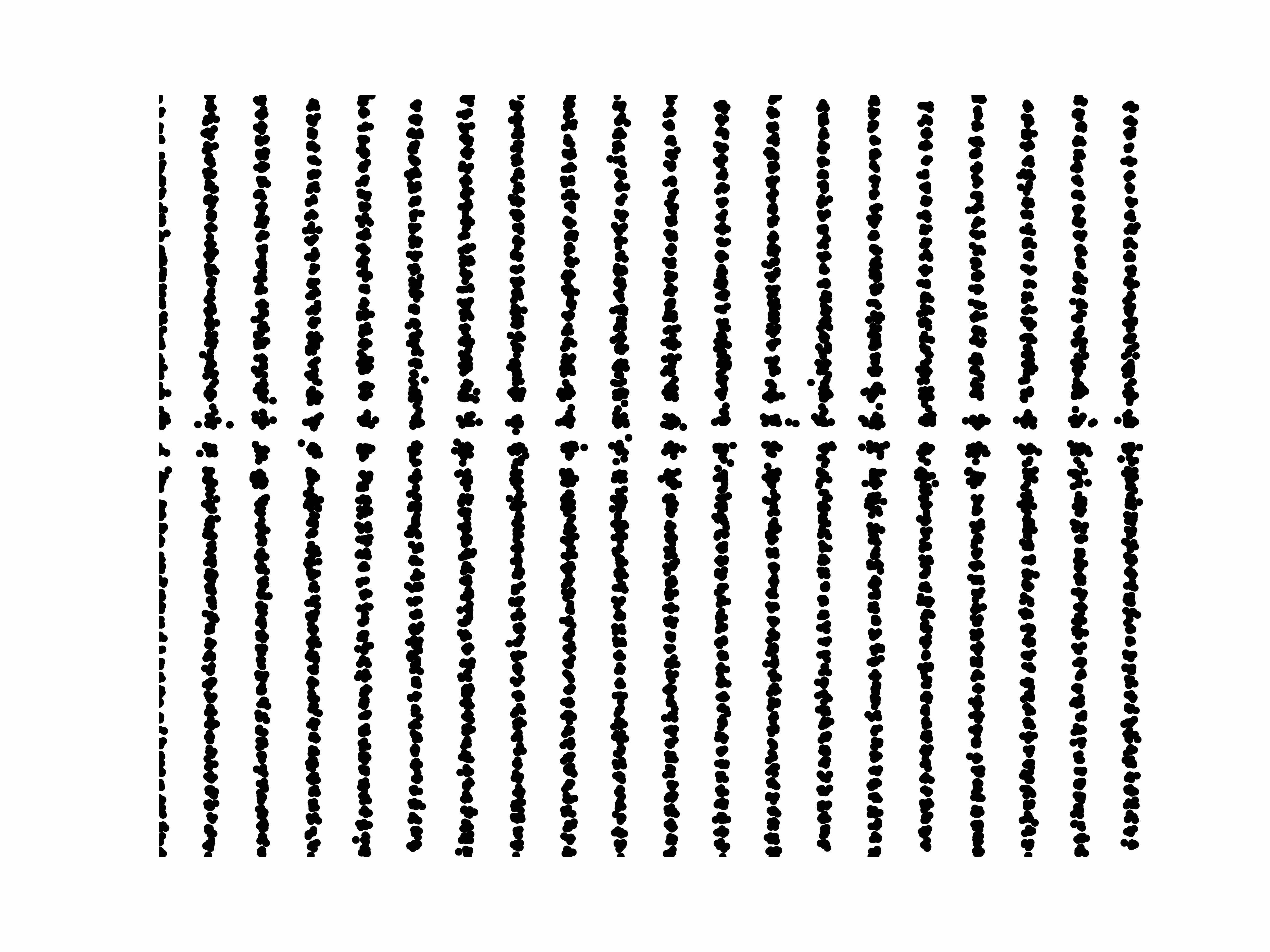}\negmedspace{}\negmedspace{}\negmedspace{}\includegraphics[clip,scale=0.033]{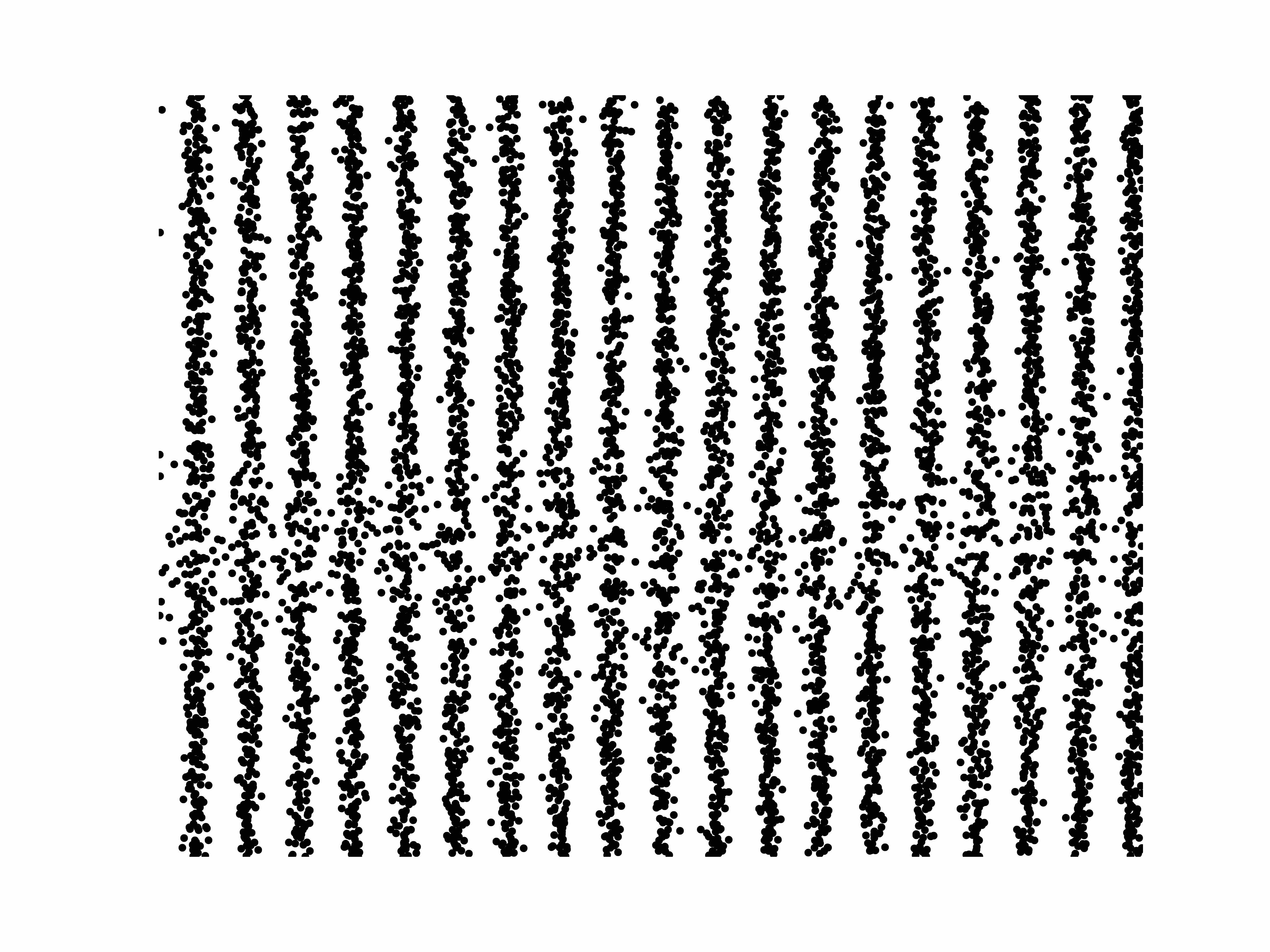}\negmedspace{}\negmedspace{}\negmedspace{}\includegraphics[clip,scale=0.033]{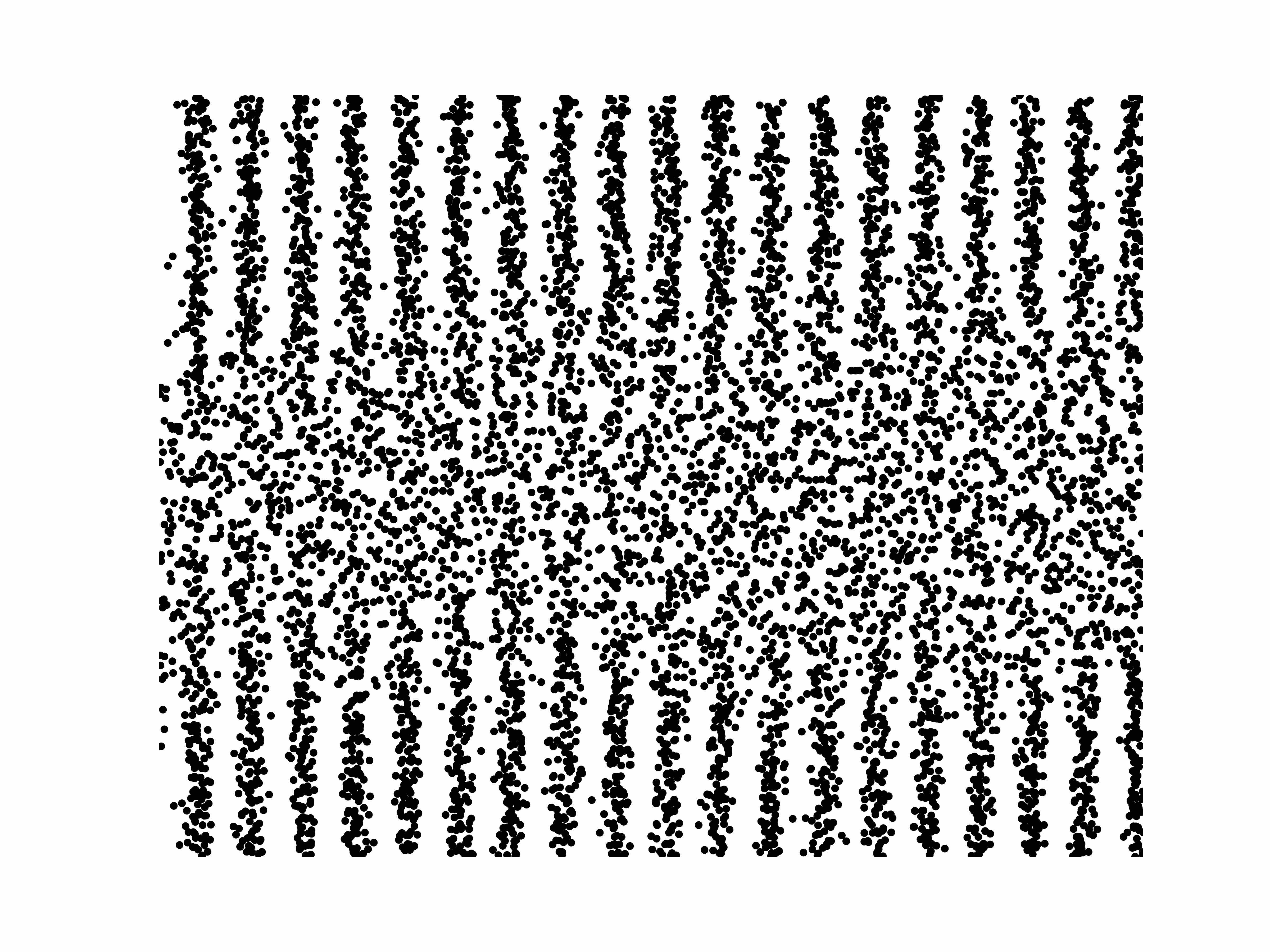}

\noindent \begin{centering}
\textbf{300 K}\qquad{}\qquad{}\qquad{}\qquad{}\qquad{}\qquad{}\textbf{1200
K}\qquad{}\qquad{}\qquad{}\qquad{}\qquad{}\textbf{1320 K}
\par\end{centering}

\protect\caption{Representative structures of the $\Sigma17$ GB in Cu at three different
temperatures. The GB plane is horizontal and the atomic positions
are projected onto the page. Top row: the tilt axis $[001]$ is normal
to the page. Lower row: the tilt axis $[001]$ is horizontal.\label{fig:Structures} }
\end{figure}

At low temperatures, direct MD calculations of $D_{gb}$ are unfeasible
and a different approach must be applied. It is assumed that ordered
GB structures can supports vacancies and interstitials and that GB
diffusion is mediated by jumps of these defects. Accordingly, a single
point defect (vacancy or interstitial) is created at various positions
within the GB core and its formation free energy is calculated using
molecular statics and the harmonic approximation to atomic vibrations
\citep{Mishin05e}. Knowing the defect free energies, equilibrium
defect concentrations are computed for different GB sites \citep{Suzuki05a,Sorensen00,Suzuki03a,Suzuki03b,Mishin05e}.
The defect is then allowed to diffuse along the GB by running MD simulations.
While the mean-square atomic displacements remain too small for computing
$D_{gb}$ from the Einstein relation, the MD trajectories can be analyzed
to identify the most typical diffusive events caused by the defect.
The absolute rates of transitions between different states of the
defect are computed using the harmonic transition-state theory \citep{Vineyard57}
with a saddle-point search by the nudged elastic band method. The
transition rates are then compiled in a rate catalog, which is fed
into kinetic Monte Carlo (KMC) simulations \citep{Sorensen00,Suzuki03a}.
Using the available information about the defect concentrations at
different GB sites, the KMC simulations permit accurate calculations
of the diffusion coefficients in different directions in the GB core.
By repeating the calculations separately for vacancies and interstitials,
their contributions to GB diffusion can be evaluated. This approach
has been successfully applied to predict GB diffusion coefficients
in a series of crystallographically different GBs in Cu, Ag, Ni and
Al \citep{Suzuki05a,Sorensen00,Suzuki03a,Suzuki03b}.

\section{GB diffusion mechanisms at low temperatures\label{sec:low_T}}

An important side product of the KMC approach described above is the
information about diffusion mechanisms of point defects in GBs. In
addition to providing the necessary ingredients for the KMC simulations,
this information presents a great value in itself. In fact, it probably
constitutes the only source of our current knowledge about low-temperature
diffusion mechanisms in metallic GBs. 

The simulations have shown that the vacancy formation energy in GBs
is on average lower than in perfect lattice but displays strong site-to-site
variations, ranging from as low as 0.1 of the lattice value to above
the lattice value. The same is true about GB self-interstitials, whose
formation energy is also on average lower than in the lattice and
shows strong site variations. The large variability of point-defect
formation energies can be linked to the existence of alternating tension
and compression regions in the GB core \citep{Suzuki03a,Suzuki03b}.
The range of interactions of point defects with GBs is limited to
about 2 to 3 lattice spacings around the GB core. An important finding
of the simulations is that the average vacancy and interstitial formation
energies in GBs are close to each other\emph{,} suggesting that both
defects are important for GB diffusion. This is drastically different
from the perfect lattice, where interstitials have a significantly
higher formation energy than vacancies.

Point defects can exist in GBs in a variety of structural forms. Both
vacancies and interstitials can be localized at certain sites or delocalized
over an extended region comprising up to 5 to 7 atoms. Some GB sites
are incapable of supporting a stable vacancy. In such cases, when
an atom is removed from a GB site to create a vacancy, the vacant
site is immediately filled by a neighboring atom. As a result, the
vacancy is formed at the neighboring site, not at the intended site.
Delocalized and unstable vacancies are common forms of GB defects,
especially in high-energy boundaries. GB interstitials can also exist
in different structural forms. They can be localized in a relatively
open space (\textquotedblleft pore\textquotedblright ) in the GB structure,
can form split dumbbell configurations similar to lattice interstitials,
or be delocalized over a region with strong atomic displacements. 

Simulations have revealed a number of GB diffusion mechanisms that
are profoundly different from the known mechanisms of lattice diffusion.
Such mechanisms often represent complex atomic rearrangements involving
a collective displacement of several atoms. Vacancies can move by
simple exchanges with a neighboring atom, as they do in the lattice,
but can also induce collective jumps of two to three atoms at a time
\citep{Suzuki03a}. Such collective jumps are caused by the existence
of unstable vacancies and have been found in many boundaries. Interstitial
atoms can migrate by hopping between interstitial positions or by
indirect jumps involving a collective displacement of several atoms.
In such collective jumps, an interstitial atom kicks out a neighboring
regular atom into an interstitial position in a neighboring structural
unit and takes its place. The displaced atom, in turn, can kick out
its neighbor so that the latter becomes an interstitial atom, etc.
Interstitial jumps involving up to four atoms moving in a concerted
manner were found in some GBs. Point defects can also induce ring
processes, in which a group of atoms implements a collective displacement
in a cyclic manner \citep{Sorensen00}. As proposed in \citep{Suzuki05a},
a prototypical diffusive event in GBs is a collective displacement
of an atomic chain (string) in which the head atom fills a relatively
open space (free volume) while the trailing atom leaves a similar
open space behind.

\section{GB diffusion mechanisms at high temperatures\label{sec:high_T}}

At high temperatures, the atomic structure of GBs becomes increasingly
disordered. Many high-angle GBs accumulate so much disorder that they
essentially turn into a liquid-like layer below the bulk melting point
$T_{m}$. In this premelting temperature range, GB diffusion rapidly
accelerates and approached the diffusivity of bulk liquid at $T\rightarrow T_{m}$
\citep{Keblinski99a,Suzuki05a,Frolov09c}. Furthermore, at premelting
temperatures the diffusion coefficients in different GBs tend to converge,
or even merge together, which suggests that GB diffusion becomes insensitive
to GB structure. Importantly, such mergers can occur below the premelting
temperature range when the GBs are still relatively ordered. Based
on this observation, it was suggested \citep{Keblinski99a,Suzuki05a}
that at high temperatures, GB diffusion is dominated by a ``universal''
mechanism similar to that in the liquid phase.

The existence of liquid-like GB diffusion remained little more than
a hypothesis for many years. Over the past decade, however, new evidence
has emerged suggesting that this hypothesis might in fact be true.
Recent atomistic investigations of stress-driven GB migration mechanisms
in Ni \citep{Zhang2006,Zhang07,Warren2009,Zhang2010} revealed a close
similarity between the atomic dynamics during the GB motion, on one
hand, and diffusion in glass-forming liquids, on the other hand. To
gain insights into GB dynamics, the authors \citep{Zhang2006,Zhang07,Warren2009,Zhang2010}
applied a set of statistical tools that had been developed for studying
the atomic dynamics in supercooled liquids and glasses \citep{Kob_97,Douglas98,Donati_1999,Zhang2013}.
One of the most interesting finding was the observation of highly
cooperative atomic motion in the form of strings of atoms. Similar
strings of collectively moving atoms had been earlier found in MD
simulations glass-forming Lennard-Jones mixtures \citep{Douglas98}
and other disordered systems. Furthermore, the string-like motion
of atoms was observed not only during GB migration but also in stationary
boundaries. This important finding suggests that the string-like motion
could be an intrinsic feature of GB dynamics and may contribute to
GB diffusion. This suggestion is extremely interesting as it provides
a closure with the earlier studies of GB diffusion at low temperatures,
where similar string-like processes were identified in atomically
ordered GBs (Sect.~\ref{sec:low_T}). Thus, cooperative atomic displacements
in the form of chains, or strings, could be one of the general diffusive
events in both ordered and disordered boundaries. In fact, this mechanism
could also operate in other types of materials interfaces, as suggested
by the recent finding of string-like motion on the surface of Ni nano-particles
\citep{Zhang2010}.

\begin{figure}
\noindent \begin{centering}
\textbf{(a)}\includegraphics[clip,scale=0.45]{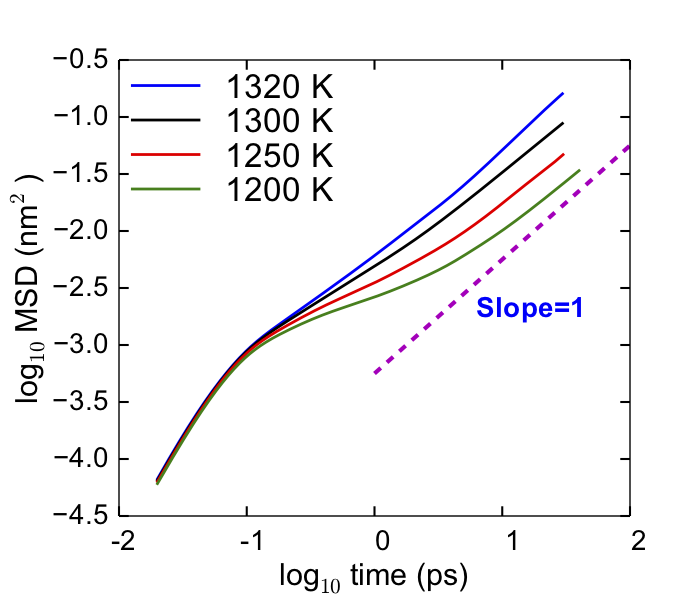}\textbf{(b)}\includegraphics[scale=0.45]{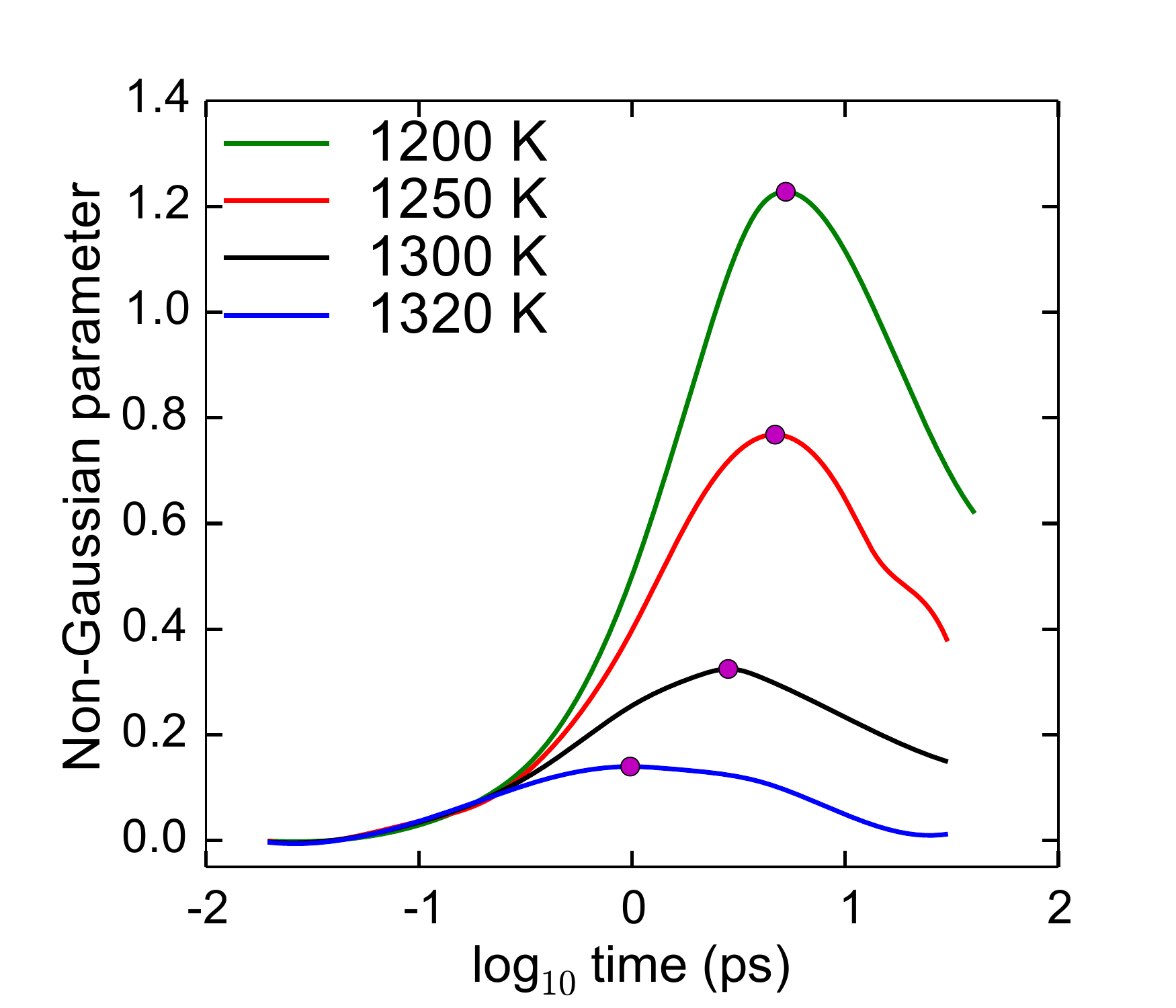}
\par\end{centering}

\protect\caption{The mean-square displacement (a) and the non-Gaussian parameter (b)
as functions of time at several temperatures. The position of the
maximum of the non-Gaussian parameter (marked by a circle) defines
the diffusion activation time $t^{*}$.\label{fig:MSN_NGP}}
\end{figure}

To further investigate the analogy between the atomic dynamics in
GBs and in supercooled liquids, we have performed a series of MD simulations
of high temperature self-diffusion in a high-angle GB in Cu, including
the premelting temperature range. As in the previous studies \citep{Zhang2006,Zhang07,Warren2009,Zhang2010},
we analyze the atomic movements by the statistical methods developed
for supercooled liquids \citep{Kob_97,Douglas98,Donati_1999,Zhang2013}.
The boundary chosen for this study is the $\Sigma17\thinspace(530)\thinspace[001]$
symmetrical tilt GB with the misorientation angle of $61.93^{\circ}$.
The ground-state structure of this boundary consists of identical
kite-shaped structural units, see Fig.~4c in \citep{Cahn06b}. Atomic
interactions were described by the EAM Cu potential \citep{Mishin01}
predicting the melting temperature of $T_{m}=1327$ K. GB diffusion
was studied at the temperatures of 1200 K, 1250 K, 1300 K and 1320
K. At each temperature, MD simulations were run in the NPT ensemble
to allow for thermal expansion of the lattice, followed by a run in
the micro-canonical (NVE) ensemble. The latter was used in order to
avoid any possible influence of a thermostat on thermal fluctuations
in the system. The statistics of atomic displacements over a given
time interval $t$ were analyzed by comparing a few hundred pairs
of MD snapshots separated by the time $t$. In each snapshot, the
GB position $y_{0}$ was determined from the peak of the potential
energy of atoms averaged over thin layers parallel to the GB plane.
All atoms within the range $y_{0}-1.5r_{0}\leq y\leq y_{0}+1.5r_{0}$
were declared GB atoms. Here, $r_{0}\approx0.25$ nm is the first-neighbor
distance in FCC Cu, which is close to the position of the first peak
in the radial distribution function in the boundary. 

\begin{figure}
\noindent \begin{centering}
\textbf{(a)}\includegraphics[clip,scale=0.45]{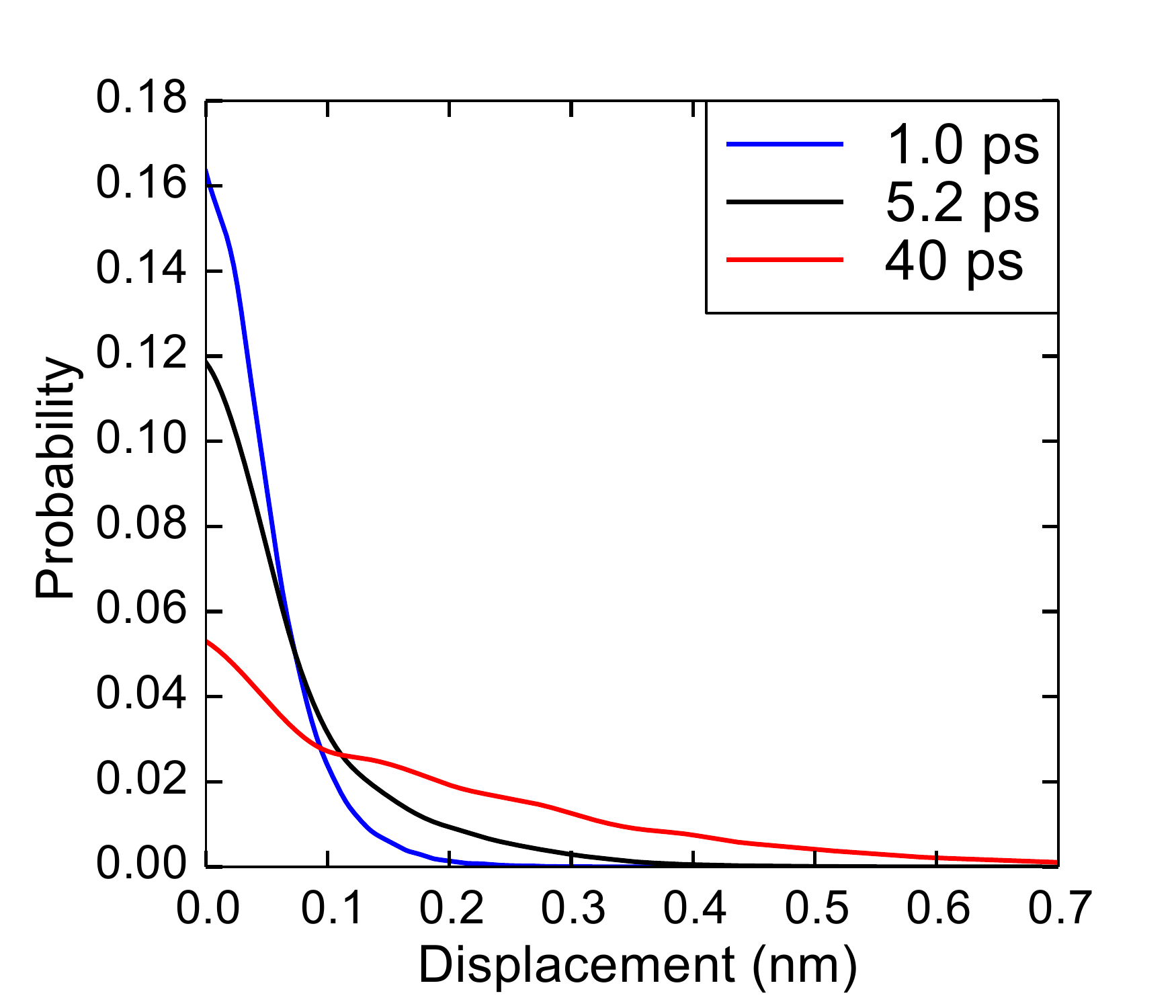}\textbf{(b)}\includegraphics[clip,scale=0.45]{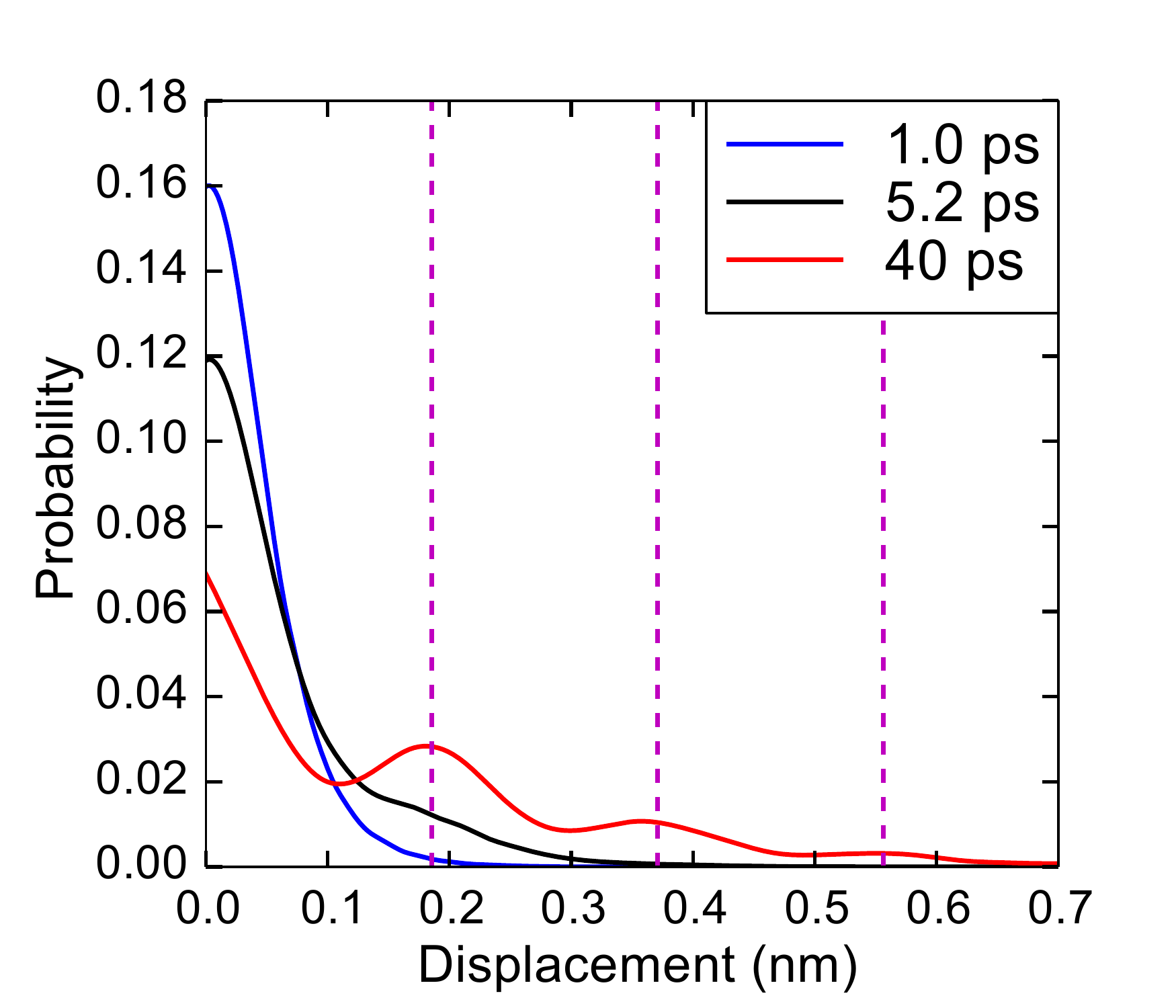}
\par\end{centering}

\protect\caption{Displacement distribution functions: (a) $W(|x|,t)$ normal and (b)
$W(|z|,t)$ parallel to the tilt axis at the temperature of 1250 K.
In (b), the vertical dashed lines mark positions of $(002)$ crystal
planes perpendicular to the tilt axis.\label{fig:Displacement-distribution}}
\end{figure}

Typical GB structures are displayed in Fig.~\ref{fig:Structures},
showing gradual accumulation of disorder with temperature and eventually
formation of a liquid-like layer 7 K below the melting point. At each
temperature, mean square displacements $\left\langle x^{2}\right\rangle $
and $\left\langle z^{2}\right\rangle $ of the GB atoms were computed
for the $x$-direction normal to the tilt axis and the $z$-direction
parallel to the tilt axis, respectively. Their average value is plotted
in Fig.~\ref{fig:MSN_NGP}a as a function of time over three orders
of magnitude (note the logarithmic scale of the plot). The plots have
the familiar $\sigma$-shape with a characteristic shoulder developing
with decreasing temperature and eventually approaching a plateau.
A similar behavior was observed in previous MD simulations of supercooled
liquids \citep{Kob_97,Douglas98,Donati_1999,Zhang2013} and GB dynamics
in Ni \citep{Zhang2006,Zhang07,Warren2009,Zhang2010}. The initial
part of the curve represents atomic vibrations and ballistic motion
of atoms implementing a diffusive jump. Once the jump is complete,
the atoms are trapped in new positions and spend some time (residence
time) waiting for a new jump. This delay in jumping explains the formation
of the shoulder. Given more time (> 1 ps), the atoms become capable
of escaping from the traps (``cages'') and can make further diffusive
jumps. On a still longer time scale, massive diffusion commences and
the plot becomes linear with the slope of 1. This signals the onset
of the full-fledged diffusive regime with the mean-square displacement
proportional to time (Einstein's relation). The time $t^{*}$ corresponding
to the middle of the shoulder and signifying the onset of GB diffusion
can be referred to as the \emph{diffusion activation time}. 

The significance of $t^{*}$ is further illustrated by the displacement
distribution plots shown in Fig.~\ref{fig:Displacement-distribution}.
The displacement distribution function $W(|x|,t)$ is the probability
that a GB atom initially located at $x=0$ will be found with a projection
$|x|$ after a time $t$, with a similar meaning for $W(|z|,t)$.
On a short time scale, the plots have a Gaussian shape and are dominated
by atomic vibrations around equilibrium. On a long time scale, the
GB atoms are capable of making multiple diffusive jumps and the plots
develop a set of local maxima corresponding to trapped positions.
Such maxima are especially pronounced in the $z$-direction due to
the relatively large spacing of the $(002)$ planes. An atom jumping
across such planes resides in a new position for some time before
making a new jump, which creates a local maximum of probability. There
is a transition time when the tail of the probability curve is just
beginning to form a local maximum, as exemplified by the curves for
$t=5.2$ ps. At this time, only a small fraction of atoms have implemented
a jump while all other atoms remain trapped. This time can be associated
with the diffusion activation time $t^{*}$. Due to the presence of
two types of atoms, those which moved and those which did not, the
probability distribution curve is significantly non-Gaussian. This
can be quantified by computing the one-dimensional non-Gaussian parameters,
\begin{equation}
p_{x}=\frac{\langle x^{4}\rangle}{3\langle x^{2}\rangle}-1,\text{\quad\quad}p_{z}=\frac{\langle z^{4}\rangle}{3\langle z^{2}\rangle}-1,\label{eq:1}
\end{equation}
for both directions. Their average, $(p_{x}+p_{z})/2$, is plotted
in Fig.~\ref{fig:MSN_NGP}b as a function of time. As expected, the
maximum of the non-Gaussian parameter closely correlates with the
middle of the shoulder in Fig.~\ref{fig:MSN_NGP}a. Therefore, we
take the position of the maximum as the definition of the diffusion
activation time. The obtained values of  $t^{*}$ are listed in Table
\ref{tab:Table1}. 

\begin{table}
\noindent \begin{centering}
\begin{tabular}{cccccc}
\hline 
$T$(K) & $t^{*}$ (ps) & $R_{c}$ (nm) & $\alpha_{c}$ & $\bar{n}$ & $d$\tabularnewline
\hline 
$1200$ & $5.26$ & $3.42$ & $79^{\circ}$ & 4.5 & 1.46\tabularnewline
$1250$ & $4.67$ & $3.59$ & $78^{\circ}$ & 5.6 & 1.49\tabularnewline
$1300$ & $2.83$ & $3.62$ & $74^{\circ}$ & 4.9 & 1.49\tabularnewline
$1320$ & $0.98$ & $3.25$ & $61^{\circ}$ & 4.8 & 1.46\tabularnewline
\hline 
\end{tabular}
\par\end{centering}

\protect\caption{Dynamic characteristics of atoms in the Cu $\Sigma17$ GB at different
temperatures: diffusion activation time $t^{*}$, correlation radius
$R_{c}$ and minimum correlation angle $\alpha_{c}$ for directional
correlations, average number of atoms in a correlated cluster $\bar{n}$,
and fractal dimension of correlated clusters $d$. \label{tab:Table1}}
\end{table}

\begin{figure}
\begin{minipage}[t]{0.43\columnwidth}%
\noindent \begin{center}
\includegraphics[clip,scale=0.41]{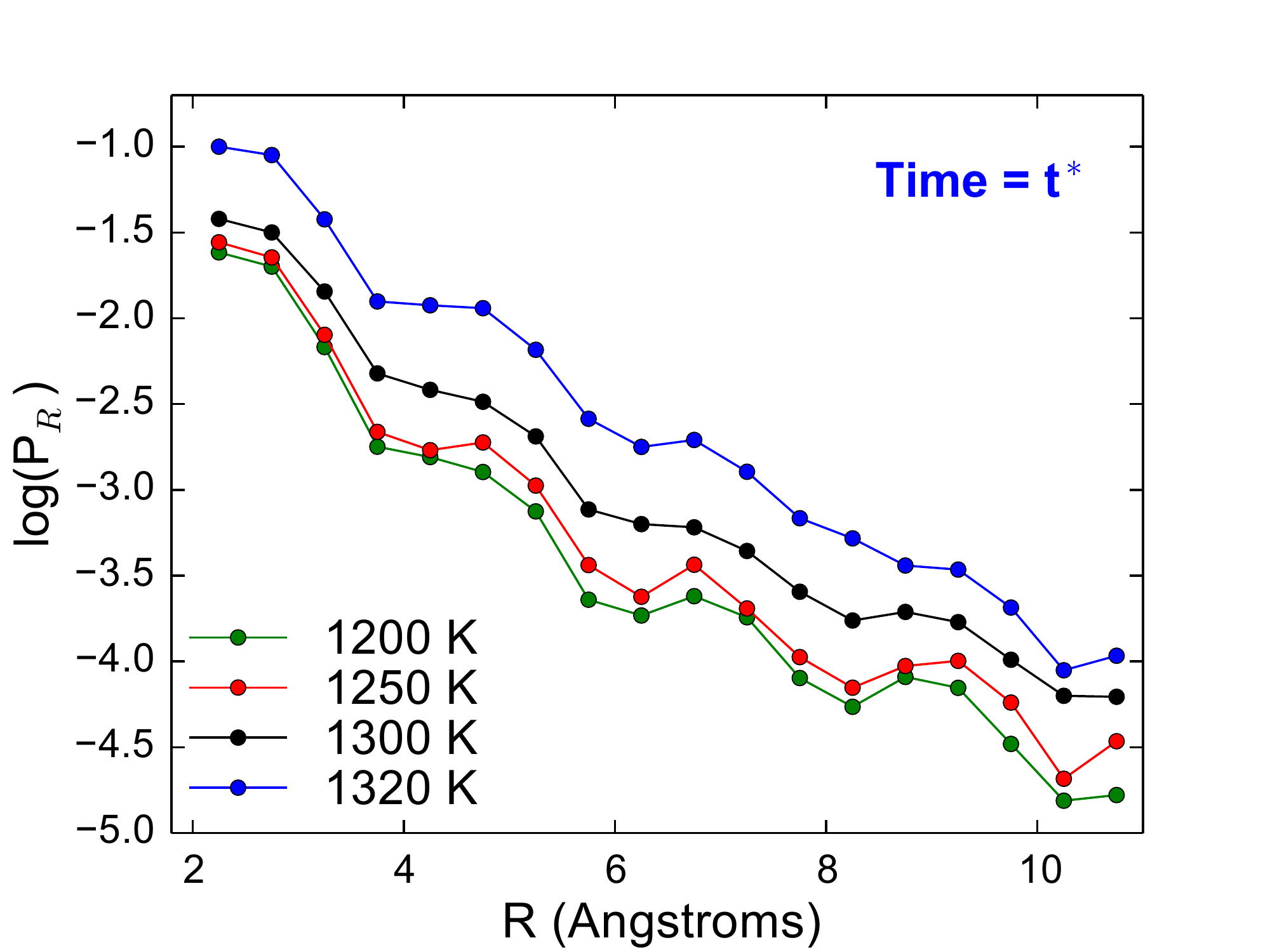}
\par\end{center}

\protect\caption{Directional correlation parameter $\mathcal{P}_{R}$ as a function
of radius $R$.\label{fig:Directional-correlation}}
\end{minipage}\hfill{}%
\begin{minipage}[t]{0.52\columnwidth}%
\noindent \begin{center}
\includegraphics[clip,scale=0.35]{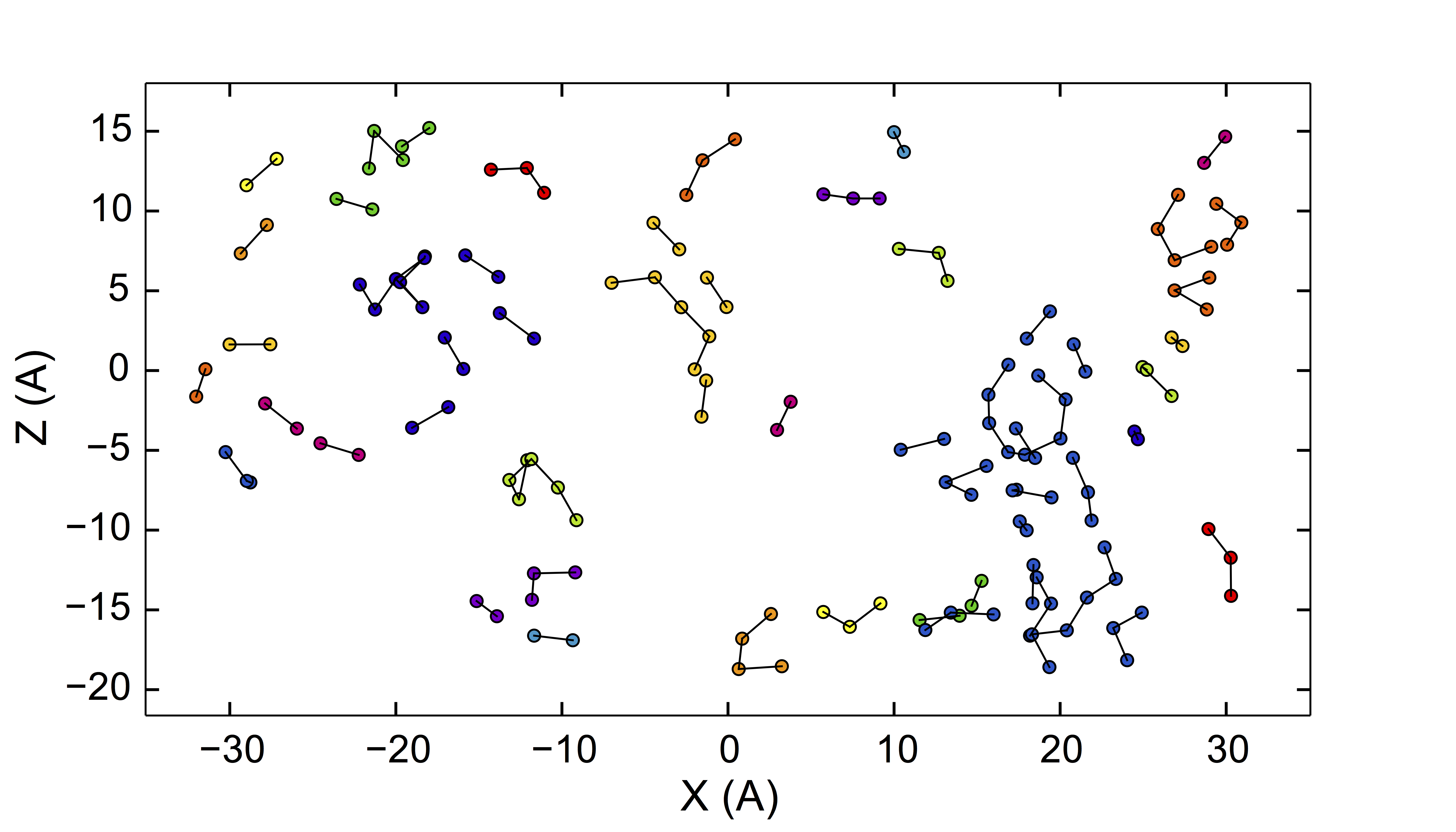}
\par\end{center}

\protect\caption{Example of correlated clusters in the Cu $\Sigma17$ boundary at the
temperature of 1250 K. Only atoms forming the clusters are shown,
with correlated pairs connected by a line for clarity. The GB plane
is parallel to the page and the tilt axis $[001]$ is vertical. \label{fig:clusters}}
\end{minipage}
\end{figure}

During the time $t^{*}$, the atoms typically make only one jump (if
at all). Thus, this time is ideal for detecting individual diffusion
events and studying their mechanisms. In all subsequent calculations,
the time will be fixed at $t^{*}$ for the respective temperature.
Several methods have been proposed for selecting the atoms which underwent
a diffusive jump (so called ``mobile'' atoms) \citep{Kob_97,Douglas98,Donati_1999}.
In this work it was found that selecting top 50\% of the GB atoms
that after the time $t^{*}$ moved a distance of at least $0.6r_{0}$
gives the most reproducible results. Examination shows that such mobile
atoms are not distributed over the GB uniformly. Instead, they show
a very clear trend to cluster into regions of high mobility, which
exist on the time scale of $t^{*}$ and later disintegrate and appear
in new locations. This reflects the well-known phenomenon of \emph{dynamic
heterogeneity} \citep{Ediger2000} which was found in supercooled
liquids and amorphous solids \citep{Kob_97,Douglas98,Donati_1999},
as well as in recent simulations of GB dynamics \citep{Zhang2006,Zhang07,Warren2009,Zhang2010}.

Several statistical measures were applied to evaluate correlations
between the displacements of the mobile atoms. One of them probes
correlations between the directions of the displacement vectors $\mathbf{d}_{i}$
of neighboring mobile atoms after the time $t^{*}$,
\begin{equation}
\mathcal{P}_{R}=\left\langle \dfrac{1}{N(N-1)}\sum_{j\neq i}\cos\left(\mathbf{d}_{i},\mathbf{d}_{j}\right)\delta\left(\mathbf{R}+\mathbf{r}_{i}(0)-\mathbf{r}_{j}(0)\right)\right\rangle .\label{eq:2}
\end{equation}
Here, the angular brackets denote the ensemble average, $N$ is the
number of mobile atoms in the boundary, and $\mathbf{r}_{i}(0)$ and
$\mathbf{r}_{j}(0)$ are positions of mobile atoms $i$ and $j$ at
time zero. The correlation parameter $\mathcal{P}_{R}$ was computed
as a function of the radius $R$. We can also define the effective
correlation angle $\alpha_{R}=\arccos\mathcal{P}_{R}$. If jump directions
are statistically independent, then $\left\langle \cos\left(\mathbf{d}_{i},\mathbf{d}_{j}\right)\right\rangle =0$
and thus $\mathcal{P}_{R}=0$ and $\alpha_{R}=90^{\circ}$ for any
$R$. The simulations give $\mathcal{P}_{R}>0$ for any $R$, proving
the existence of directional correlations: neighbors of an atom tend
to move in approximately the same direction as the atom. By analogy
with other relaxation processes we expect an exponential decay of
$\mathcal{P}_{R}$ with $R$,
\begin{equation}
\mathcal{P}_{R}=\mathcal{P}_{max}\exp\left(-R/R_{c}\right).\label{eq:3}
\end{equation}
Fig.~\ref{fig:Directional-correlation} shows that the computed $\mathcal{P}_{R}(R)$
curves approximately follow Eq.(\ref{eq:3}). The obtained correlation
radius $R_{c}$ is approximately 0.35 nm at all temperatures tested
(Table \ref{tab:Table1}). The value of $\alpha_{c}=\arccos\mathcal{P}_{max}$
can be interpreted as the smallest correlation angle, which in our
case varies between $60^{\circ}$ and $80^{\circ}$.

\begin{figure}
\noindent \begin{centering}
\includegraphics[clip,scale=0.45]{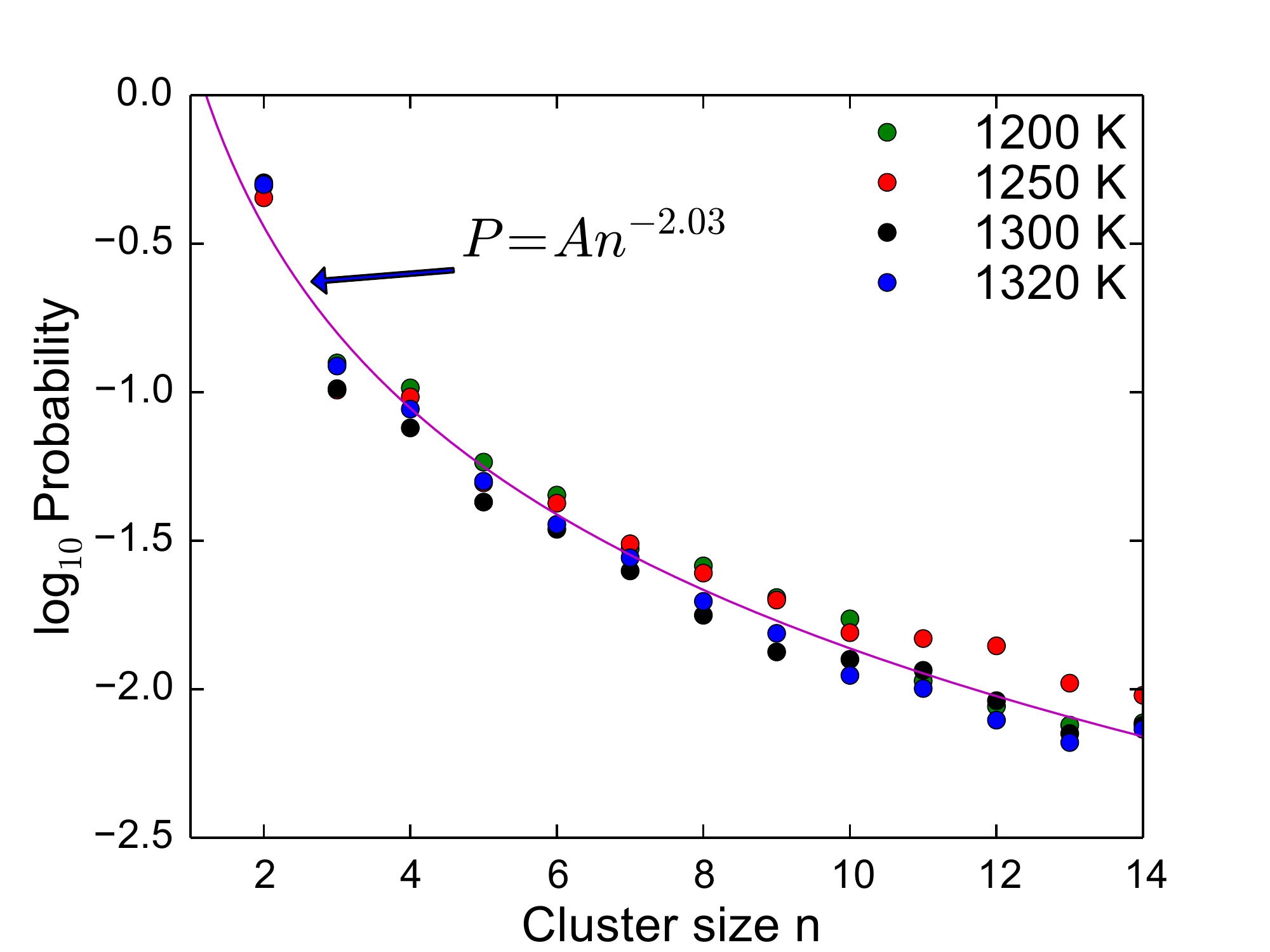}\negthickspace{}\negthickspace{}\negthickspace{}\negthickspace{}\includegraphics[clip,scale=0.45]{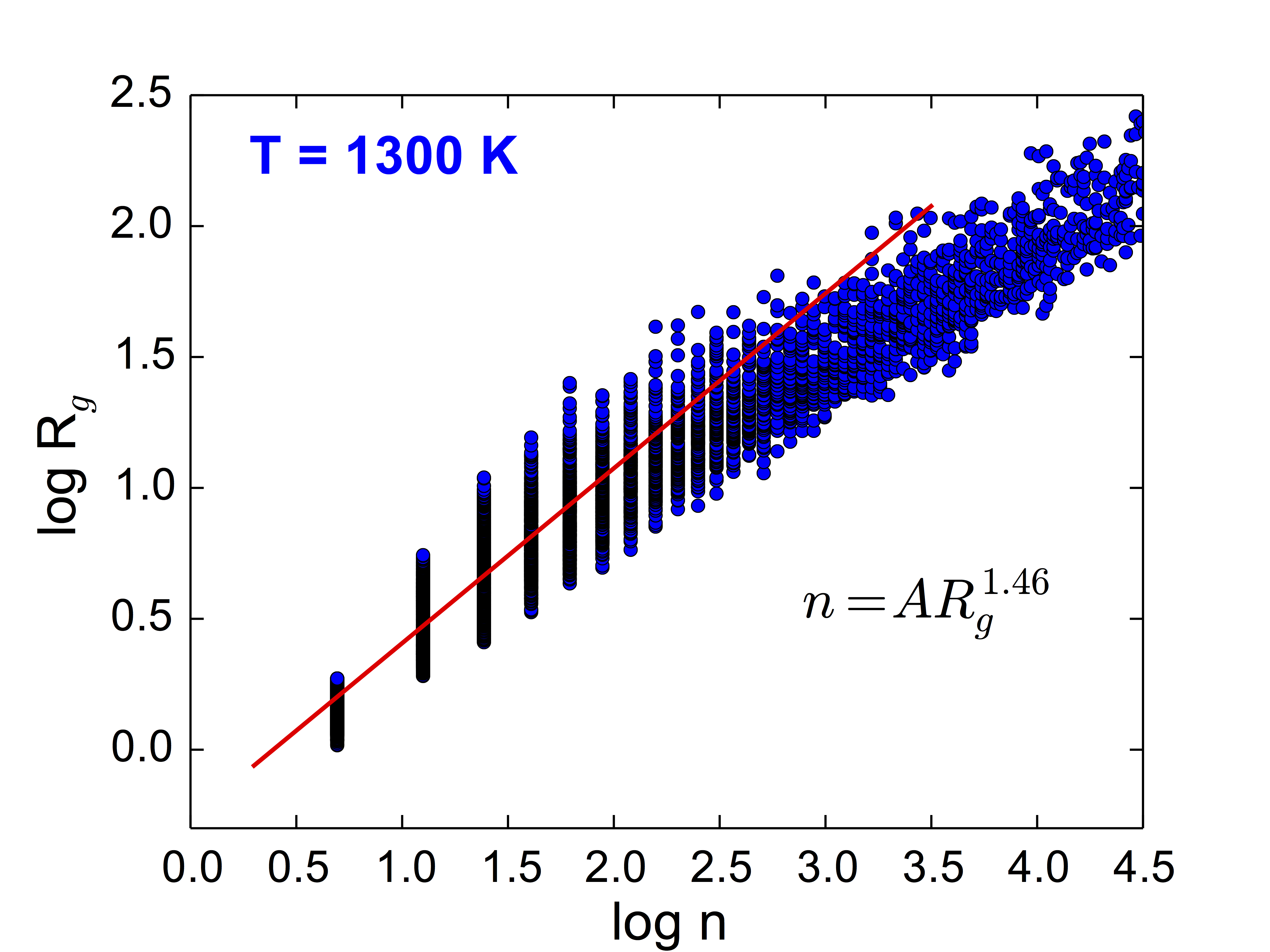}
\par\end{centering}

\noindent \begin{centering}
\textbf{(a)\qquad{}\qquad{}\qquad{}\qquad{}\qquad{}\qquad{}(b)}
\par\end{centering}

\protect\caption{Analysis of correlated clusters in the Cu $\Sigma17$ boundary. (a)
Probability distribution of the number of atoms in a cluster. The
line shows a fit by a power law $P\propto n^{-\beta}$. (b) Example
of a log-log plot of the number of atoms in a cluster versus its gyration
radius $R_{g}$. The slope of the linear fit is the fractal dimension
of the clusters.\label{fig:Analysis-of-clusters}}
\end{figure}

Next, we define ``correlated pairs'' as pairs of mobile GB atoms separated
by a distance less than $1.05r_{0}=0.2625$ nm at both $t=0$ and
$t=t^{*}$; in other words, neighbors remain neighbors after the time
period $t^{*}$. An important observation is that such correlated
pairs tend to form \emph{clusters} composed of two, three or more
pairs as illustrated in Fig.~\ref{fig:clusters}. The probability
distribution $P(n)$ of the number $n$ of atoms in a cluster does
not depend on temperature within the statistical scatter of the points
(Fig.~\ref{fig:Analysis-of-clusters}a). At variance to the previous
findings \citep{Kob_97,Douglas98,Donati_1999,Zhang2006,Zhang07,Warren2009,Zhang2010},
$P(n)$ is not exponential. Rather, it can be fitted by a power law
$P\propto n^{-\beta}$ with $\beta$ close to 2. This relation calls
for a theoretical explanation in the future. The average value of
$n$ is around 5 (Table \ref{tab:Table1}). 

Visual inspection of the clusters indicates that many of them have
the shape of a string in agreement with previous work \citep{Kob_97,Douglas98,Donati_1999,Zhang2006,Zhang07,Warren2009,Zhang2010}.
To quantify this trend, the gyration radius $R_{g}$ was computed
for each cluster and log-log plots of $R_{g}$ versus $n$ were used
to estimate the fractal dimension $d$ of the clusters as defined
by the relation $R_{g}\propto n^{d}$ (see example in Fig.~\ref{fig:Analysis-of-clusters}b).
The obtained values of $d$ are remarkably independent of temperature
and are close to $1.47$ (Table \ref{tab:Table1}). That $d$ is smaller
than 3 (compact clusters) or even 2 (e.g., disk-shape objects) confirms
that many clusters are indeed one-dimensional objects such as strings. 

\begin{figure}
\begin{minipage}[b]{0.47\columnwidth}%
\includegraphics[clip,scale=0.45]{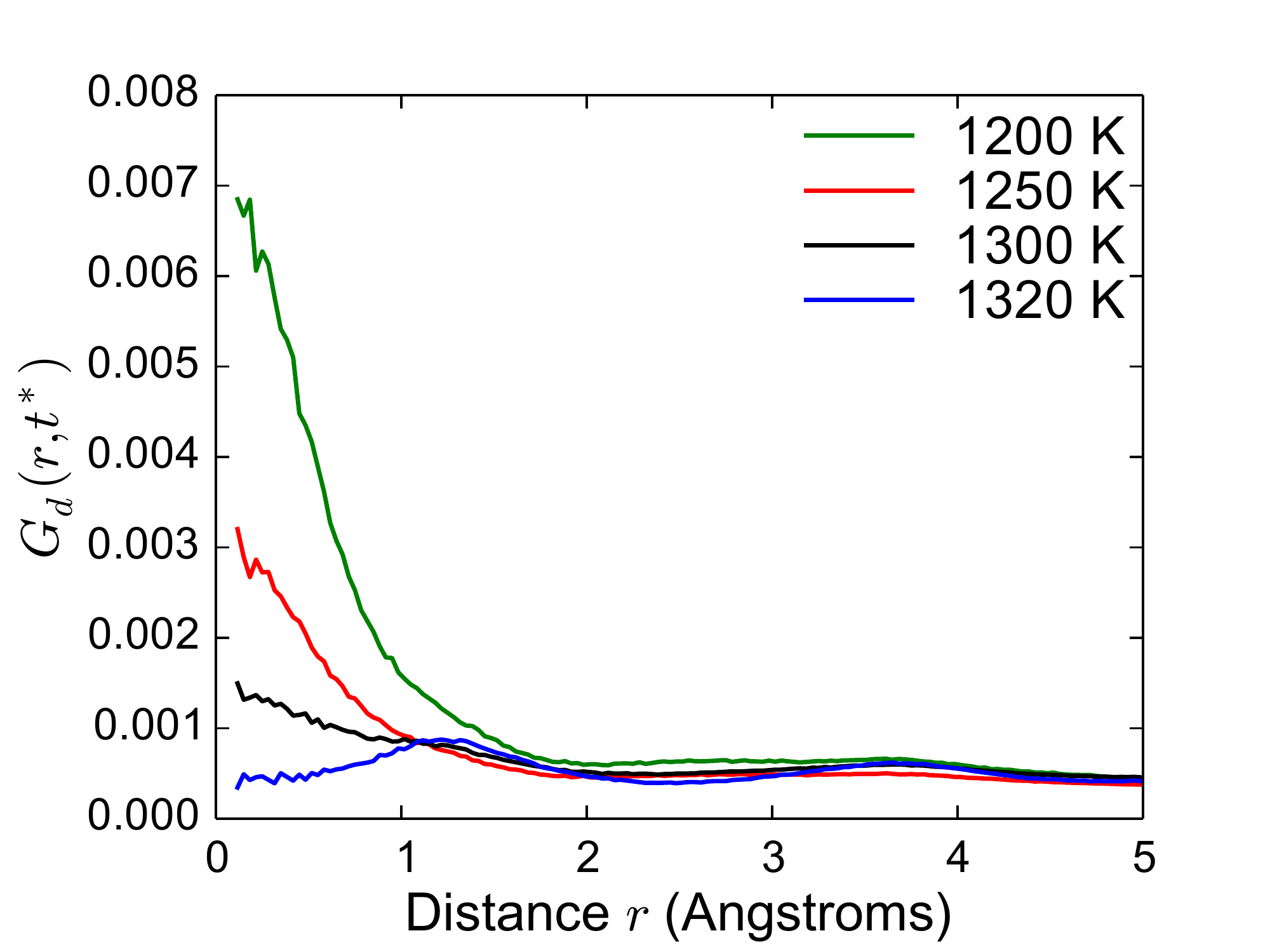}%
\end{minipage}\hfill{}%
\begin{minipage}[t]{0.47\columnwidth}%
\vspace{-4cm}
\protect\caption{The distinct part $G_{d}(r,t^{*})$ of the van Hove correlation function
for mobile GB atoms as a function of radius $r$. Note the pea near
$r=0$ growing with decreasing temperature. \label{fig:van Hove}}
\end{minipage}
\end{figure}

Another aspect of atomic dynamics in GBs is the hopping character
of the correlated displacements. This aspect can be analyzed using
the direct-space van Hove correlation function \citep{Hove1954}
\begin{equation}
G\left(\mathbf{r},t\right)=\left\langle \dfrac{1}{N}\sum_{i}\sum_{j}\delta\left(\mathbf{r}+\mathbf{r}_{j}(0)-\mathbf{r}_{i}(t)\right)\right\rangle =G_{s}\left(\mathbf{r},t\right)+G_{d}\left(\mathbf{r},t\right),\label{eq:4}
\end{equation}
which is naturally separated into a self-part
\begin{equation}
G_{s}\left(\mathbf{r},t\right)=\left\langle \dfrac{1}{N}\sum_{i}\delta\left(\mathbf{r}+\mathbf{r}_{i}(0)-\mathbf{r}_{i}(t)\right)\right\rangle \label{eq:5}
\end{equation}
related to the displacement distribution function $W$ discussed above,
and a distinct part
\begin{equation}
G_{d}\left(\mathbf{r},t\right)=\left\langle \dfrac{1}{N}\sum_{i\neq j}\delta\left(\mathbf{r}+\mathbf{r}_{j}(0)-\mathbf{r}_{i}(t)\right)\right\rangle .\label{eq:6}
\end{equation}
The distinct part characterizes the distribution of atoms $j$ at
a time $t$ around a site that was occupied by a particle $i\neq j$
at time $t=0$. Fig.~\ref{fig:van Hove} shows radial distributions
$G_{d}(r,t^{*})$ for mobile GB atoms at different temperatures. As
temperature decreases, a prominent peak grows near zero $r$. This
peak indicates that there is a higher than random probability that
when an atom undergoes a diffusive transition to a different location,
its place is taken by another mobile atoms on the time scale of $t^{*}$.
That this ``substitutional'' character of diffusion is especially
pronounces at lower temperatures is consistent with the increase in
the atomic order of the boundary with decreasing temperature. The
peak disappears at 1320 K when the GB turns into a layer of liquid
with a small supercooling (7 K).

\section{Summary}

Recent computer simulations have shed light on atomic-level diffusion
mechanisms in GBs and possibly other interfaces. At low temperatures
when GB structure is well-ordered, diffusion is mediated by point
defects - vacancies and interstitials. However, by contrast to diffusion
in the perfect lattice, not all GB sites can support a point defect,
and when they can, the defect structure is often delocalized over
an extended region. GB atoms can move by either individual jumps or
by collective displacements involving several atoms moving simultaneously.
Such collective displacements frequently occur in the form of chains,
or strings, in which the head atom fills a relatively open space and
the trailing atom leaves a similar open space behind. 

At high temperatures, GB structure becomes disordered and in some
cases forms a liquid-like layer. The notion of individually moving
point defects loses its significance. Instead, GB diffusion can be
described by statistical methods developed for analyzing atomic dynamics
in supercooled liquids, glasses, and other disordered systems of strongly
interacting particles. Recent applications of these methods to GBs
reported in the literature, as well as in the present paper, reveal
a remarkable similarity between GB diffusion at high temperatures
and diffusion in supercooled liquids. A picture which emerges from
these studies is that of GB atoms diffusing by collective displacements
of whole groups (clusters) comprising two to five, and possibly more,
atoms. Many of these clusters are one-dimensional objects similar
to the strings found at low temperatures. When the boundary maintains
some degree of atomic order, the correlated displacements of the clusters
have a discrete character, i.e., occur by collective hopping from
one equilibrium position to another. When the boundary forms a liquid
layer, the collective motion of atoms remains but has a more continuous
character. In the future, it would be extremely interesting to extend
such studies to GB diffusion in alloys with strong segregation, especially
in systems where the liquid phase undergoes a glass transition. 

This work was supported by the National Science Foundation, the Metals
and Metallic Nanostructures Program, under Award \# 1308667. 


\end{document}